\documentclass[preprint,aps,superscriptaddress]{revtex4}%
\usepackage{amsfonts}
\usepackage{amsmath}
\usepackage{amssymb}
\usepackage[dvips]{graphicx}%
\begin{document}
\title{A Cold Atomic Fermi Gas with a Spatially Modulated Interaction}
\author{Hao Fu}
\affiliation{MCTP, FOCUS Center, and Physics Department, University of Michigan, Ann Arbor,
Michigan 48109-1040}
\author{Alberto G. Rojo}
\affiliation{Department of Physics, Oakland University, Rochester, MI 48309}

\pacs{03.75.Hh, 05.30.Jp}

\begin{abstract}
We study an ultra-cold atomic Fermi Gas with the atom-atom interaction
modulated periodically in space. A novel ground state with cooper pairs
occupying non-zero center of mass momenta is found. Such a state is closely
related to the state proposed by Fulde, Ferrell, Larkin, and
Ovchinnikov(FFLO). The resultant single particle excitations with momenta
along the direction of the modulation shows multiple-gap structures. Such a
system can be realized in experiment with a spatially modulated Feshbach
resonance. Experimental signatures of such a state are discussed.

\end{abstract}
\date{October, 30, 2006}
\maketitle

In several recent experiments, polarized Fermi gases have been
realized\cite{pfe}. These progresses drive a new wave of studies of the so
called FFLO state\cite{ff}\cite{lo}.\ Such a state is of great interest
because of its implications in many systems, including color superconducting
quark matter, dense nuclear matter, and superconducting systems in solid
states\cite{mvs}. The observation of such a state, however, is still elusive
because it exists only in a very narrow parameter regime in cold atomic gases.
In addition, it is not completely clear whether such a state favors occupation
of multiple\cite{lo} or single\cite{ff} center of mass (COM) states. In this
paper, we propose to study a ground state closely related to the FFLO state.
At zero temperature in a homogeneous system, it is the only ground state of
the system and the study of such a state should shed light on the study of the
FFLO state.

We consider a fermion gas with the atom-atom interaction varying in space. For
an ultra cold atomic gas, the effective atom-atom interaction is, $V\left(
\mathbf{\rho}\right)  =4\pi a_{s}\hbar^{2}\delta\left(  \mathbf{\rho}\right)
/m$ where $\mathbf{\rho}$ is relative coordinate between two particles, $m$ is
the mass of the atom, and $a_{s}$ is the s-wave scattering length. Near a
Feshbach resonance, the scattering length can be described by \cite{aeff}
$a_{eff}=a_{bg}-\frac{m}{4\pi\hbar^{2}v_{0}}\left\vert g_{0}\right\vert ^{2}$,
where $a_{bg}$ is the background scattering length, and the second term is the
contribution from the Feshbach resonance nearby. The constants $g_{0}$ and
$v_{0}$ are the coupling strength and energy detuning between the scattering
channel and the molecular channel, respectively. In magnetic Feshbach
resonances\cite{aeff}\cite{mfr}, the detuning $v_{0}$ can be controlled by an
external magnetic field. In an optical Feshbach resonance\cite{oft}\cite{ofe}
$v_{0}$ can be tuned by laser frequencies, and the free bound coupling $g_{0}$
can be tuned by the laser intensity. A spatially dependent scattering length
can be achieved by applying spatially varying external fields. This gives an
effective interaction, $V\left(  \mathbf{r},\mathbf{\rho}\right)  =4\pi
a_{s}\left(  \mathbf{r}\right)  \hbar^{2}\delta\left(  \mathbf{\rho}\right)
/m,$ that depends on both the relative coordinate and the center of mass (COM)
coordinate $\mathbf{r}$. This direct substitution of the scattering length
locally requires that the scattering length to vary only in a scale much
larger than the local scattering length. In particular, we consider the
simplest form of such a interaction,
\begin{equation}
V\left(  \mathbf{r},\mathbf{\rho}\right)  =\left[  g_{0}+g\cos\left(
\mathbf{k}_{0}\cdot\mathbf{r}\right)  \right]  \delta\left(  \mathbf{\rho
}\right)  \label{inter}%
\end{equation}
The constant $g_{0}$ term, with $g_{0}<g<0,$ is added to guarantee that the
system stays in the BCS side of the resonance. Such a system, which to the
best of our knowledge have not been studied before, is only feasible with
recent advances in atomic physics. A relatively obvious phenomenon that arises
from such a spatially modulated interaction would be the modulation of the
atomic density. Atoms are prone to congregate in locations with maximum
attractions. A much more interesting phenomenon resides in the pairing of
atoms. In a typical BCS theory, the ground state wave function consists of
Cooper pairs with their COM momenta equal to zero. \ The interaction we
propose (\ref{inter}) creates coherences between pairs with COM momenta
differing by $\mathbf{k}_{0}$ in order to lower the free energy of the system.
The ground state thus includes Cooper pairs with zero, as well as
$\pm\mathbf{k}_{0}$, $\pm2\mathbf{k}_{0}$ COM momenta. However, we find that
the probability of occupying higher COM momenta is low due to disadvantage in
kinetic energies and the constant attractive interaction $g_{0}$. Our ground
state is closely related to the FFLO state discussed in the
literature\cite{ff}\cite{lo}. In the FFLO state, due to the mismatch of the
chemical potential for spin up and spin down particles, particles can pair
with non zero COM momenta to lower the energy. Before proceeding with our
analysis, we emphasize that although our ground state is similar to the FFLO
one, the spatial symmetry is spontaneously broken in the FFLO case, while in
our case the spatial symmetry is broken by an external field.

We work in a grand canonical ensemble and use a single channel Hamiltonian
which is appropriate for the case of a broad Feshbach resonances,
\begin{align}
K  &  =\int d\mathbf{r}\sum_{a}\psi_{\alpha}^{\dag}\left(  \mathbf{r}\right)
\left(  -\frac{1}{2m}\bigtriangledown^{2}-\mu\right)  \psi_{\alpha}\left(
\mathbf{r}\right) \nonumber\\
&  +\frac{1}{2}\int d\mathbf{r}\sum_{\alpha,\beta}\tilde{V}\left(
\mathbf{r}\right)  \psi_{\alpha}^{\dag}\left(  \mathbf{r}\right)  \psi_{\beta
}^{\dag}\left(  \mathbf{r}\right)  \psi_{\beta}\left(  \mathbf{r}\right)
\psi_{a}\left(  \mathbf{r}\right)  \label{tp}%
\end{align}
where $\mu$ is the chemical potential, and $\psi_{\alpha}\left(
\mathbf{r}\right)  $, $\psi_{\alpha}^{\dag}\left(  \mathbf{r}\right)  $ are
the fermionic annihilation and creation operators with spin index $\alpha$ at
position $\mathbf{r}$. In the interaction part, the relative coordinate is
already integrated out. The $\tilde{V}\left(  \mathbf{r}\right)  $ is the
standard renormalized interaction\cite{renorm} from our bare interaction
(\ref{inter}). We solve this system in a mean field level at zero temperature
using the Bogoliubov de Gennes(BdG) transformation\cite{bdg}%
\begin{align}
\left(  \varepsilon_{k}-\mu\right)  u_{\mathbf{k}}+\tilde{U}_{\mathbf{k}%
^{\prime}}u_{\mathbf{k}-\mathbf{k}^{\prime}}+\Delta_{\mathbf{k}^{\prime}%
}v_{\mathbf{k}-\mathbf{k}^{\prime}}  &  =Ev_{\mathbf{k}}\label{bdg1}\\
-\left(  \varepsilon_{k}-\mu\right)  v_{\mathbf{k}}-\tilde{U}_{\mathbf{k}%
^{\prime}}v_{\mathbf{k}-\mathbf{k}^{\prime}}+\Delta_{\mathbf{k}^{\prime}%
}^{\ast}u_{\mathbf{k}+\mathbf{k}^{\prime}}  &  =Eu_{\mathbf{k}}\label{bdg2}\\
-\frac{1}{V}v_{\mathbf{k}^{\prime}}^{\left(  n\right)  }v_{\mathbf{k+k}%
^{\prime}-\mathbf{k}^{\prime\prime}}^{\left(  n\right)  }\tilde{g}%
_{\mathbf{k}^{\prime\prime}}  &  =U_{\mathbf{k}}\label{bdg3}\\
\frac{1}{V}v_{\mathbf{k}^{\prime}}^{\left(  n\right)  }u_{\mathbf{k+k}%
^{\prime}-\mathbf{k}^{\prime\prime}}^{\left(  n\right)  }\tilde{g}%
_{\mathbf{k}^{\prime\prime}}  &  =\Delta_{\mathbf{k}}\label{bdg4}\\
\frac{2}{V}v_{\mathbf{k}}^{\left(  n\right)  }v_{\mathbf{k}}^{\left(
n\right)  }  &  =\bar{n} \label{bdg5}%
\end{align}
A summation convention is used for any momentum that appears twice on the
left. We take the gap as $\Delta_{\beta\alpha}\left(  \mathbf{r}\right)
=-g\left(  \mathbf{r}\right)  \left\langle \psi_{\beta}\left(  \mathbf{r}%
\right)  \psi_{a}\left(  \mathbf{r}\right)  \right\rangle $ and Hatree
potential as $U_{\alpha\alpha}\left(  \mathbf{r}\right)  =g\left(
\mathbf{r}\right)  \left\langle \psi_{\alpha}^{\dagger}\left(  \mathbf{r}%
\right)  \psi_{\alpha}\left(  \mathbf{r}\right)  \right\rangle $. The spin
index can be neglected with no confusion due to the symmetry between spin up
and down. In the following we take $\mathbf{k}_{0}$ along $z$ direction. With
this choice, $\Delta\left(  \mathbf{r}\right)  ,$ $U\left(  \mathbf{r}\right)
$ are functions of $z$ only. In momentum space, we can solve the single
particle problem in a set of subspaces corresponding to constant $k_{x}$ and
$k_{y}.$ The Fourier transforms of $u\left(  \mathbf{r}\right)  ,v\left(
\mathbf{r}\right)  ,\tilde{V}\left(  \mathbf{r}\right)  _{,}\Delta\left(
\mathbf{r}\right)  $ and $U\left(  \mathbf{r}\right)  $ are defined as
$u_{\mathbf{k}},v_{\mathbf{k}},g_{\mathbf{k},}\Delta_{\mathbf{k}}$ and
$U_{\mathbf{k}}$ respectively. We can take $u\left(  \mathbf{r}\right)
,v\left(  \mathbf{r}\right)  $ and $\Delta_{\mathbf{k}}$to be real without
loss of generality. Since we are dealing with a problem invariant under
reflection $\mathbf{r}\rightarrow-\mathbf{r}$, we can also take $a_{\mathbf{k}%
},b_{\mathbf{k}},\Delta_{\mathbf{k}}$ and $U_{\mathbf{k}}$ to be real.
$\varepsilon_{\mathbf{k}}=\hbar^{2}k^{2}/2m$ is the kinetic energy of a
particle with wave vector $\mathbf{k},$ and $V$ is the quantization volume. We
note that near a Feshbach resonance the chemical potential, $\mu,$ need to be
self-consistently determined by an average density of the Fermi gas $\bar{n}$
in equation (\ref{bdg5}).

\begin{figure}
[ptb]
\begin{center}
\includegraphics[
height=5.3483in, width=6.5039in
]%
{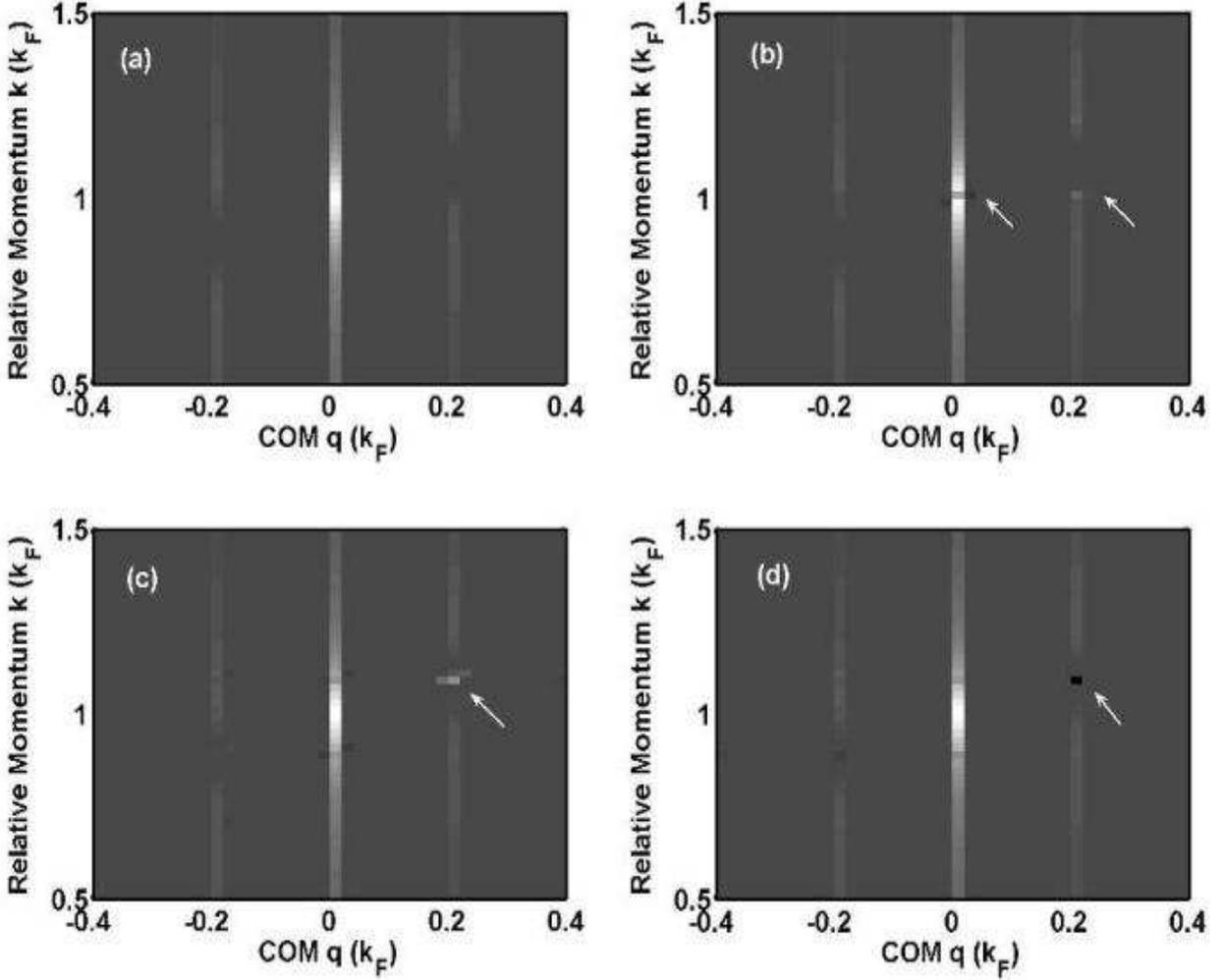}%
\caption{Pairing amplitudes $a_{k_{z}\mathbf{\uparrow}}a_{q-k_{z}%
\mathbf{\downarrow}}$ for $k_{x}=k_{y}=0$. Plot (a) corresponds to the ground
state, (b) to the first excited state, (c) to states right before the second
gap, and (d) tp the state right after the second gap. Only momenta close to
the fermi surface are shown. The arrows are used as guide to the eye to pair
amplitudes where significant changes taken place. We have used a units with
fermi momentum and fermi energy to be one. The numerical values for the
parameters are: $g_{0}=20,$ $g=15,$ and $k_{0}=0.2$. The self-consistant gaps
and chemical potential are found to be, $\Delta_{0}=.13$, $\Delta_{\pm k_{0}%
}=.06$ and $\mu=0.85.$}%
\label{guv}%
\end{center}
\end{figure}

We solve the BdG equations self-consistently by numerical iteration. The
self-consistent gap is found to have only three components in momentum space,
namely, $k_{z}=0,\pm k_{0}.$ Due to the form of interaction we
chose(\ref{inter}), the zero momentum component gap is always larger than the
component of momenta $\pm k_{0}$ and therefore the gap is nonzero everywhere
in real space. This is different from the pairing gap discussed in the FFLO
papers\cite{ff}\cite{lo}, where the gap has only $k_{0}$ or $\pm k_{0}$
components. The pairing of the many body ground state is given by
$\left\langle a_{\mathbf{k\uparrow}}a_{\mathbf{q}-\mathbf{k\downarrow}%
}\right\rangle $. Note that in the $k_{x}-k_{y}$ plane, the pairing occurs
with opposite momenta. In the $z$ direction, however, there are non zero COM
$q$ values. We study the pairing of atoms with their $k_{x}=k_{y}=0$ without
loss of generality. It can be shown that $\left\langle a_{k_{z}%
\mathbf{\uparrow}}a_{q-k_{z}\mathbf{\downarrow}}\right\rangle =\sum
_{n}u_{k_{z}}^{\left(  n\right)  }v_{k_{z}-q}^{\left(  n\right)  }$ and the
numerical result is presented in figure(\ref{guv}a).%

We find that only states with COM momenta $\mathbf{q}=0,\pm\mathbf{k}_{0}$ are
occupied. The higher COM momentum states with integer multiples of
$\mathbf{k}_{0}$ have negligible wight. Given the magnitude of the coupling
between different COM states components, it is surprising that atom pairs do
not occupy states with COM momenta being higher harmonics of $k_{0}$. Actually
this result can be explained by the nonlinear property of the BdG equations.
Atom pairs tend to occupy as few COM states as possible because the
interaction energy is proportional to the square of $\left\langle
a_{k_{z}\mathbf{\uparrow}}a_{q-k_{z}\mathbf{\downarrow}}\right\rangle $. By
occupying only these three COM states, the pairing energy is maximized in
magnitude. The pairing occurs mostly for atoms near the Fermi surface. For a
particular $k_{z}$, the pairing amplitudes for the COM $q=\pm k_{0}$ are
generally not equivalent, i.e. $\left\langle a_{k_{z}\mathbf{\uparrow}%
}a_{k_{0}-k_{z}\mathbf{\downarrow}}\right\rangle \neq\left\langle
a_{k_{z}\mathbf{\uparrow}}a_{-k_{0}-k_{z}\mathbf{\downarrow}}\right\rangle .$
This is because for $k_{z}\neq0,$ either $k_{0}-k_{z}$ or $-k_{0}-k_{z}$ is
closer to the Fermi surface. Since pairs can occupy several COM states
coherently, one particular atom, say $a_{\mathbf{k}\uparrow}^{\dag},$ can form
pairs simultaneously with several particles, say $a_{-\mathbf{k}\downarrow},$
$a_{\mathbf{k}_{0}-\mathbf{k}\downarrow},$and $a_{-\mathbf{k}_{0}%
-\mathbf{k}\downarrow}$. Right at the Fermi surface, pairings with opposite
momenta dominate, and therefore the amplitudes $\left\langle a_{\mathbf{k}%
_{F}\uparrow}a_{\mathbf{k}_{0}-\mathbf{k}_{F}\downarrow}\right\rangle $ and
$\left\langle a_{\mathbf{k}_{F}\uparrow}a_{-\mathbf{k}_{0}-\mathbf{k}%
_{F}\downarrow}\right\rangle $ are small. Slightly away from the Fermi
surface, the pairing with opposite COM momenta is not as strong and one starts
to see pairings with nonzero COM. This feature of the ground state pairing is
closely related to the single particle excitation spectrum discussed below.

\begin{figure}
[ptb]
\begin{center}
\includegraphics[
height=5.3483in, width=6.5039in
]%
{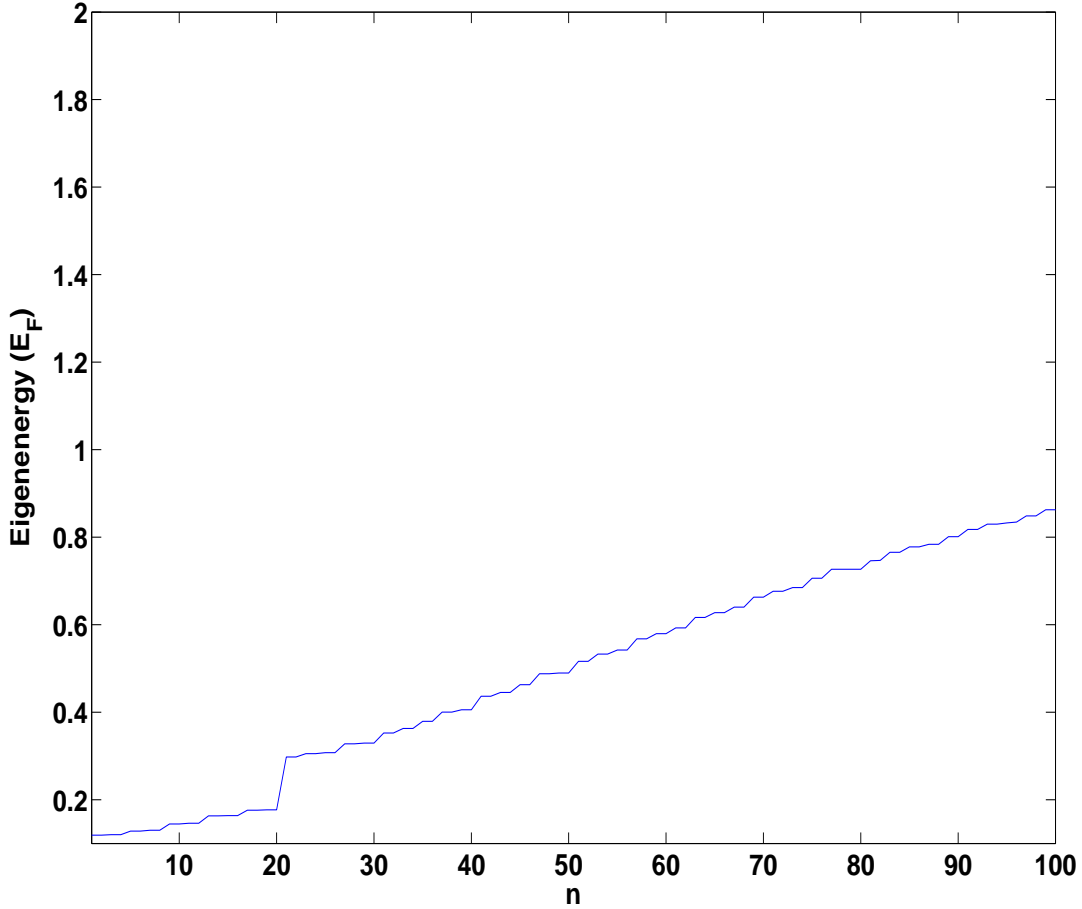}%
\caption{Plot of the single particle excitation energy. Here we only look at
the eigen--excitations with $k_{x}=k_{y}=0.$ The $n$ labels the eigenstates.
We have used same numerical values for relevent parameters as figure(\ref{guv}%
).}%
\label{pen}%
\end{center}
\end{figure}

The eigenenergies that we found by solving (\ref{bdg1})(\ref{bdg2}) are the
single particle excitation spectrum of the many body state. The excitation is
always gapped as opposed to the state studied by FFLO\cite{ff}\cite{lo}. The
quantity $\Delta\left(  \mathbf{r}\right)  $ is positive everywhere in real
space with its minimum at $\Delta_{0}-\Delta_{k_{0}}$. This minimum is the
lower bound of the excitation gap. Above the gap, the excitation spectrum is
continuous due to the continuous distribution of $k_{x}$ and $k_{y}$. For a
particular set of $k_{x}$ and $k_{y}$, for example $k_{x}=k_{y}=0,$ additional
gaps emerge due to the periodic modulation of the interaction along $z$
direction [see Fig(\ref{pen})].%

The modulation therefore induces band gap structures in quasi-particle
excitations. More interestingly, we can understand the excitation spectrum in
a pair breaking picture. The pairing amplitudes in an excited state
$\left\vert n\right\rangle $ are, $\left\langle n\left\vert a_{k\uparrow
}a_{q-k\downarrow}\right\vert n\right\rangle =\left\langle 0\left\vert
a_{k\uparrow}a_{q-k\downarrow}\right\vert 0\right\rangle -u_{k}^{\left(
n\right)  }v_{k-q}^{\left(  n\right)  }$. The pairing amplitudes for relevant
states are shown in Fig (\ref{guv}b, \ref{guv}c, \ref{guv}d). The pair
breaking happens simultaneously around both $-k_{F}$ and $k_{F}$. The energy
lowered by this superposition is a small effect, and in the following we focus
on one side of the Fermi surface. For the lowest excited state $\left\vert
1\right\rangle $, the change in the pairing amplitudes occurs right at the
Fermi surface. The amplitude $a_{\mathbf{k}_{F}\uparrow}a_{-\mathbf{k}%
_{F}\downarrow}$ is almost completely destroyed in state $\left\vert
1\right\rangle $. One the other hand, recall that in the ground state, there
are no pairing amplitudes for $a_{\mathbf{k}_{F}\uparrow}a_{\mathbf{k}%
_{0}-\mathbf{k}_{F}\downarrow}$ and $a_{\mathbf{k}_{F}\uparrow}a_{-\mathbf{k}%
_{0}-\mathbf{k}_{F}\downarrow},$ due to the dominant pairing of $a_{\mathbf{k}%
_{F}\uparrow}a_{-\mathbf{k}_{F}\downarrow}$. Now that the $a_{\mathbf{k}%
_{F}\uparrow}a_{-\mathbf{k}_{F}\downarrow}$ pair is broken, the particle
$a_{k_{F}\uparrow}^{\dag}$ is free to pair with particles $a_{\pm
\mathbf{k}_{0}-\mathbf{k}_{F}\downarrow}^{\dag}$ and this decrease the energy
by a amount of $\Delta_{k_{0}\text{.}}$ Since the excitation is close to the
Fermi surface, the kinetic energy contribution can be neglected. Furthermore,
the Hartree term is almost the same for the ground and excited states and it
can be neglected too. Therefore the emergence of gaps can be attribute to the
pairing energy only, namely $\sum_{k}\left[  2\Delta_{0}u_{k}^{\left(
n\right)  }v_{k}^{\left(  n\right)  }+2\Delta_{k_{0}}u_{\pm k_{0}+k}%
v_{k}\right]  .$ The gap energy can then be estimated to be $0.09$, close to
the observed value $0.10$. The excitation energy increases continuously as the
excitation gets further and further away from the fermi surface. The states
with energies right below and above the second gap have sudden flips of the
signs for the pairing amplitude $a_{k^{\prime}\uparrow}a_{\pm k_{0}-k^{\prime
}\downarrow}$. Here the $k^{\prime}$ are the particular wavevectors where
pairing amplitudes $a_{k^{\prime}\uparrow}a_{\pm k_{0}-k^{\prime}\downarrow}$
undergo significant changes with respect to the ground state. We can repeat
the analysis above and estimate the energy cost for the sign flipping to be
$0.09,$ close to the observed value $0.11$. Therefore, the phase flips of the
pairs with nonzero COM momenta give rise to the second gap. Similar
observations can be made to locate gaps at higher energies. However, as the
kinetic energy increases, the gap structure gets more and more obscure.
Actually, with parameters used in the calculation, only two gaps are observed.
We would like to emphasis that, after summation over the contribution from
different $k_{x}$ and $k_{y},$ the additional gap we discussed above, is not a
true second gap in the excitation spectrum. However, it nevertheless induces a
sharp change of density of states for energies close to the gap. Such effects,
arising from the rich structures of the pair wavefunction, can be observed in experiments.

The parameters we use in the calculation are based on recent $^{6}Li$
experiments \cite{ferexp}\cite{cropro}. We take the fermi energy
$E_{F}=\left(  3\pi^{2}\bar{n}\right)  ^{1/3}$ $\sim3\mu k$ and momentum
$k_{F}$ to be the unit energy and wave vector, respectively. In this units,
the background scattering length $a_{bg}\sim-0.5$ \ and the modulation of the
interaction (\ref{inter}) $k_{0}\sim.2$. Here we estimate $k_{0}$ to be the
order of laser wavevector. It is smaller than $k_{F}$ and this choice
automatically satisfies the condition $k_{0}\ll\left\vert 2\pi/a\right\vert $,
necessary for the validity of equation (\ref{inter}). The constant coupling
strength is about $g_{0}\sim-20$. The coupling between different COM $g$ is
chosen to be $15,$ smaller in magnitude than $g_{0}$. This coupling can be
realized in experiment both by external magnetic and optical fields. However,
it is not easy to generate spatial variations of the magnetic field in a scale
as small as the sample size. Optical Feshbach resonance, as demonstrated in
the experiment \cite{ofe}, is a more promising way to generate the spatial
modulation of the interaction. By coupling the incident channel of two atom
scattering to the molecular excited state close channel\cite{oft}, one can
induce the periodical spatial modulation by an optical lattice. The laser
field frequency needs to be chosen to be far away from the resonance of a
single atom to minimize the induced single particle effects.

\begin{figure}
[ptb]
\begin{center}
\includegraphics[
height=5.3483in, width=6.5039in
]%
{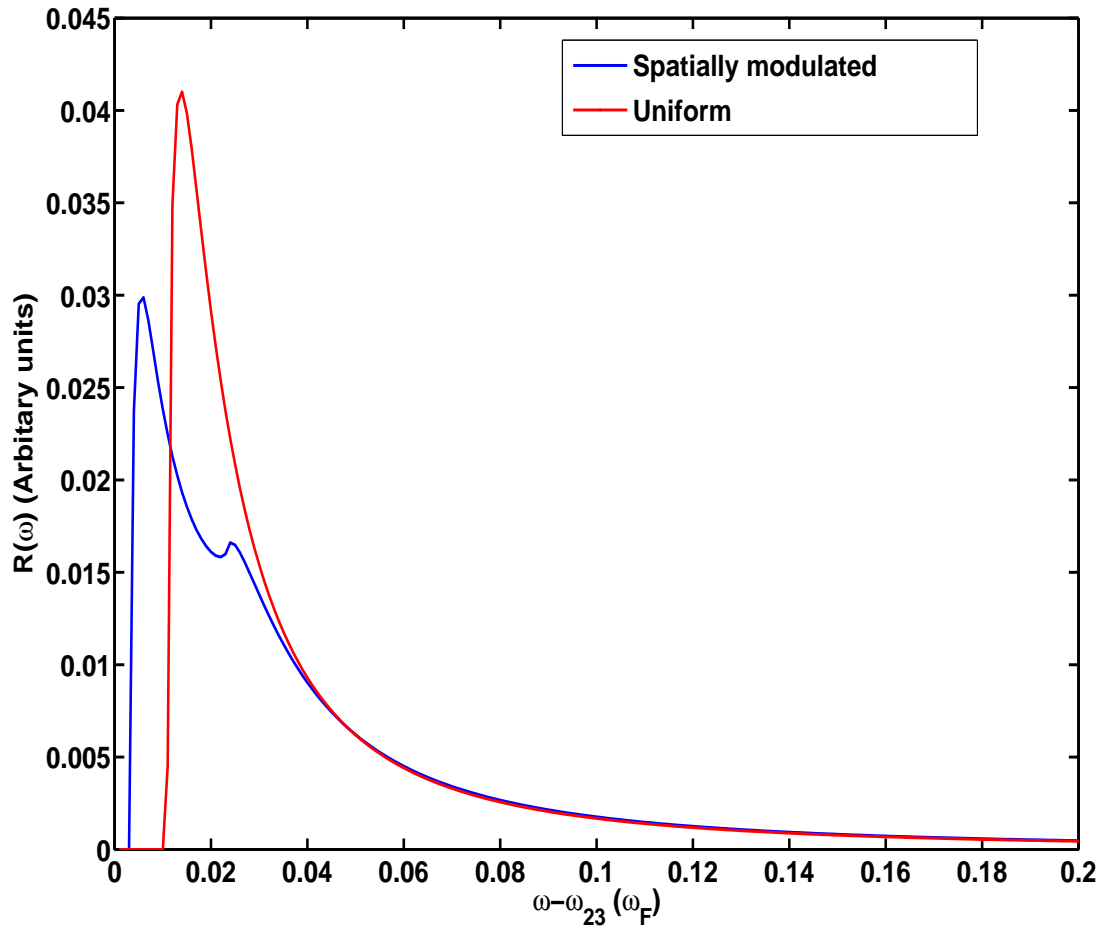}%
\caption{(color online) The RF spectrum is plotted for a homogenously
interacting system and our spatially modulated interacting system. Notice that
there is an extra peak in the RF singal for the system with a spatially
modulated interaction. We have again used the numerical values as
Figure(\ref{guv}) for various parameters. }%
\label{rfs}%
\end{center}
\end{figure}

The observation of this novel pairing in the ground state involves
measurements of the COM momenta of pairs. This can be done by pair-wise
projecting Cooper pairs into molecules by a fast sweep from BCS side of the
resonance to the BEC side. After this procedure, the molecules of the
condensate should coherently occupy momentum states $\mathbf{q}=0$,
$\mathbf{q}=\pm\mathbf{k}_{0}$, which can be measured by a time of flight
image of molecules. Such measurements have already been used in several
BEC-BCS cross over experiments\cite{cropro}. Another way of observing the
pairing correlation, recently demonstrated experimentally\cite{noise}, is to
measure the shot noise correlations. For the excited states properties, one
needs to measure the radio frequency (RF) spectrum. RF spectra have been
proposed\cite{rft} and recently used to measure the fermi gas pairing
gap\cite{rfe}. In experiments, two of the hyperfine levels of atoms,
$\left\vert 1\right\rangle ,\left\vert 2\right\rangle $ are identified as the
spin up and spin down states. A probing RF field couples one of the hyperfine
levels, say $\left\vert 2\right\rangle ,$ to a third hyperfine level
$\left\vert 3\right\rangle $. The transition rate can be evaluated directly
from our BdG solution of the quasiparticle excitations. However, Such
calculation turns out to be numerically challenging. This requires to
discretize the momentum space in a very small grid. Here, we adopt a Local
Density Approximation(LDA) approach. The calculation should be valid for
$k_{0}\ll k_{F}$, which is exactly the situation we are interested in. In LDA,
the RF spectrum is calculated locally and the final result is a summation of
local contributions \cite{rft}, $R\left(  \omega\right)  =\frac{\pi}{\hbar
}\frac{\Omega^{2}}{V}\int d^{3}rD\left(  \varepsilon_{k}\right)  \frac
{\Delta^{2}\left(  r\right)  }{\hbar^{2}\omega^{2}}\Theta\left[
\varepsilon_{k}\left(  \omega\right)  \right]  .$ Note that the local gap and
chemical potential are given by our self-consistent BdG solutions. Here
$\omega$ is the detuning of the laser frequency from the frequency difference
in $\left\vert 2\right\rangle $ and $\left\vert 3\right\rangle ,$ and $\Omega$
is the coupling between them. $D\left(  \varepsilon_{k}\right)  $ is the free
particle density of states and it is proportional to $\sqrt{\varepsilon_{k}}$.
In side the step function, the $\varepsilon_{k}\left(  \omega\right)
=\frac{\omega^{2}-\Delta\left(  r\right)  ^{2}}{2\omega}+\mu\left(  r\right)
$ should be greater than zero, which gives the threshold of the excitation.
The chemical potential of state $\left\vert 3\right\rangle $ is taken to be
zero and the RF spectrum is shown in figure (\ref{rfs}).%

We note two features of the spectrum. First, the threshold is shifted to a
lower energy comparing with the uniform case. This corresponds to a local gap
$\Delta\left(  r\right)  =\Delta_{0}-\Delta_{k_{0}}$, which is consistent with
the observation that the lowest excitation states correspond to break a COM
zero pair and increase the non-zero COM pairing. Second, there is an extra
peak at the higher energy side of the spectrum, which corresponds to a local
gap $\Delta\left(  r\right)  =\Delta_{0}+\Delta_{k_{0}}.$ This is consistent
with our analysis that the second gap originates from sigh flips of the
non-zero COM pairs and the magnitude is proportional to the $\Delta_{k_{0}}$ only.

In conclusion, we studied a fermi gas with spatially modulated interaction in
a mean field level at zero temperature. We also discussed its experimental
realization and detection. Such a state has a periodical modulation of the
order parameter similar to the FFLO states. Even though we considered the
spatially varying interaction along only one direction, our analysis can be
easily generalized to a system where the interaction is modulated in three
directions. In this case the second gaps along three directions can overlap
and produce true additional gaps of the quasiparticle excitation. This should
produce more pronounce signals in the RF spectrum.

\bigskip HF would like to thank P. R. Berman for discussions on the modulation
of scattering length and thank L.-M. Duan and W. Yi for several useful
comments. This work is done under financial support of\ NSF0244841 and FOCUS
Center grant. A.G.R acknowledges support from the Research Corporation,
Cottrell College Science Award.

\end{document}